\documentstyle[aps,prl,epsfig]{revtex}
\def\narrowtext{} \tighten
\twocolumn

\def\La40{La$_{1.2}$Sr$_{1.8}$Mn$_{2}$O$_{7}$}

\begin{document}
\draft
\title{Charge melting and polaron collapse in
La$_{1.2}$Sr$_{1.8}$Mn$_{2}$O$_{7}$}
\author{L. Vasiliu-Doloc,$^{1,2,}$\cite{NIU} S. Rosenkranz,$^3$ R.
Osborn,$^3$ S. K.  Sinha,$^3$ J. W. Lynn,$^{1,2}$ J. Mesot,$^3$
O. H. Seeck,$^3$ G. Preosti,$^{3,4}$ A. J. Fedro,$^{4}$ and
J. F. Mitchell$^3$}
\address{$^1$NIST Center for Neutron Research, National Institute of
Standards and Technology, Gaithersburg, Maryland 20899}
\address{$^2$Department of Physics, University of Maryland, College
Park, MD 20742}
\address{$^3$Argonne National Laboratory, Argonne, Illinois 60439}
\address{$^4$Department of Physics, Northern Illinois University,
DeKalb,
IL 60115}

\address{%
\begin{minipage}[t]{6.0in}
\begin{abstract}
X-ray and neutron scattering measurements directly demonstrate the
existence of polarons in the paramagnetic phase of optimally-doped colossal
magnetoresistive oxides. The polarons exhibit short-range correlations
that grow with decreasing temperature, but disappear abruptly at the
ferromagnetic transition because of the sudden charge delocalization.
The ``melting" of the charge ordering as we cool through $T_C$ occurs with
the collapse of the quasi-static polaron scattering, and provides important
new insights into the relation of polarons to colossal magnetoresistance.
\typeout{polish abstract}
\end{abstract}
\end{minipage} }

\maketitle
\narrowtext

\newpage Manganese oxides have attracted tremendous interest
because they exhibit colossal magnetoresistance (CMR) - a dramatic
increase in the electrical conductivity when they order ferromagnetically. 
The basic relationship between ferromagnetism and conductivity in doped 
manganese oxides has been understood in terms of the double-exchange 
mechanism \cite{zener,dagotto},
where an itinerant $e_{g}$ electron hops between Mn$^{4+}$ ions,
providing both the ferromagnetic exchange and electrical conduction. In
addition, an important aspect of the physics of manganese oxides is the
unusually strong coupling among spin, charge, and lattice degrees of
freedom \cite{dagotto,millis}. These couplings can be tuned by varying
the electronic doping, electronic bandwidth, and disorder, giving rise to a
complex phase diagram in which structural, magnetic, and transport
properties are intimately intertwined. 
The charge-ordered phases represent one of the most intriguing results
of balancing these couplings, and have been observed at low temperature in
insulating, antiferromagnetically ordered manganites, 
but are incompatible with double exchange-mediated ferromagnetism 
seen in optimally-doped CMR systems.

In comparison to the cubic manganites such as La$_{1-x}$A$_{x}$MnO$_{3}$
(A=Sr, Ca, Ba), the two-layer Ruddlesden-Popper compounds
La$_{2-2x}$Sr$_{1+2x}$Mn$_{2}$O$_{7}$\cite{moritomo}, where $x$ is the 
nominal hole concentration, are advantageous to study because the reduced
dimensionality strongly enhances the spin and charge fluctuations. The 
crystal structure is
body-centered tetragonal (space group $I4/mmm$) \cite{mitchell} with 
$a\simeq 3.87$ \AA\ and $c\simeq 20.15$ \AA , and consists of MnO$_{2}$
bilayers separated by (La,Sr)O sheets. In the intermediate doping regime
($0.32 \leq x < 0.42$), the ground state is a ferromagnetic
metal, and the magnetoresistance is found to be strongly enhanced near
the combined metal-insulator and Curie transition at $T_{C}$ (112 K
for the $x$=0.4 system of present interest\cite{prl}). The present 
results reveal diffuse scattering associated with lattice distortions 
around localized charges, i.e. polarons, in the paramagnetic phase.
The formation of lattice polarons above the ferromagnetic transition 
temperature $T_{C}$ has been inferred from a variety of measurements 
\cite{billinge}, but detailed observation via diffuse
x-ray or neutron scattering in single crystals has been lacking until
now \cite{shimomura}. Through such measurements, we have 
observed the collapse of quasi-static polaron
scattering when the metallic, ferromagnetic state is entered.
Furthermore, we present evidence of the growth of relatively well
developed short-range polaron correlations in the paramagnetic phase of
this optimally-doped CMR material. However, the development of
long-range charge ordering is preempted by the delocalization of the 
polarons themselves at $T_C$.

The measurements were performed on a single-crystal of the double-layer
compound La$_{1.2}$Sr$_{1.8}$Mn$_{2}$O$_{7}$, with dimensions 
$6\times 4\times 1$ mm$^{3}$, cleaved from a boule that was grown using 
the floating-zone technique \cite{mitchell}. The x-ray data were taken 
on the 1-ID-C diffractometer at the Advanced Photon Source, mostly using 
a high-energy beam of 36 keV to provide enough penetration in 
transmission geometry. Additional measurements were taken in reflection 
geometry with 21 keV. The neutron measurements were performed on the BT-2 
triple-axis spectrometer at the NIST research reactor, using both 
unpolarized (with either energy integration or energy analysis) and 
polarized neutron beams
with an incident energy of 13.7 meV. For the measurements under magnetic
field at BT-2, we employed a superconducting solenoid to provide fields
up to 9 T applied in the $ab$ plane. A wide range of reciprocal space was
explored, including the ($h$0$l$) and ($hhl$) planes.

A polaron consists of a localized charge with its associated
lattice distortion field, which gives rise to diffuse scattering 
around the Bragg peaks, known as Huang scattering. 
Figure 1(a) shows a contour plot of the diffuse x-ray scattering in 
the ($h$0$l$) plane around the (0, 0, 8), (0, 0, 10) 
and (0, 0, 12) reflections \cite{magnrod}. A similar

\begin{figure}[tbp]
\vspace*{-2.25cm}
\centerline{\epsfig{file=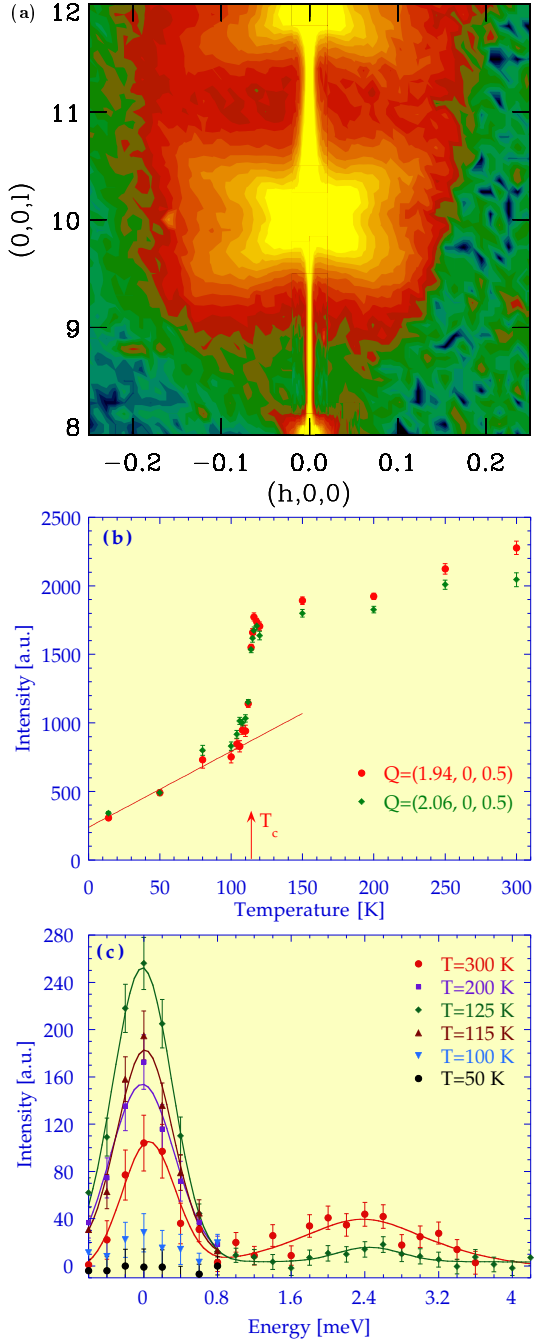,width=14.9truecm}}

\vspace*{-1.0cm}
\caption{(a) Contour plot showing the lobe-shaped pattern of diffuse
x-ray scattering at $T$=300 K around the (0, 0, 8), (0, 0, 10) and
(0, 0, 12) reflections. (b) Observed temperature-dependence of the two
$l>0$ lobes of diffuse x-ray scattering around (2, 0, 0). The straight 
line at low $T$ is the estimated phonon contribution (thermal diffuse
scattering), while the abrupt jump near $T_C$ is due
to the formation of polarons. (c) Neutron energy scans for several
different temperatures at a wave vector Q=(2.05, 0, 0.25), which is on 
one of the lobes of diffuse scattering around the (2, 0, 0)
Bragg reflection. The excitation at $\sim$ 2.4 meV is an
acoustic phonon. A flat background of 29 counts plus an elastic
incoherent peak of 89 counts, measured at 10 K, have been subtracted 
from these data.}
\end{figure}

\noindent
anisotropic pattern of diffuse scattering was observed around (2, 0, 0).
This scattering has a strong temperature dependence, with a dramatic
response at $T_C$, as illustrated in Fig. 1(b). The almost linear
temperature dependence of the diffuse scattering below $T_C$ in Fig.
1(b) suggests that phonons dominate in this temperature
regime \cite{isaacs}, but the sudden change at $T_C$ cannot be due to
conventional acoustic phonons. This is confirmed by neutron energy scans
such as shown in Fig. 1(c), which reveal both quasi-elastic and inelastic
(phonon) contributions. The phonon mode at about 2.4 meV is well
separated from the quasi-elastic scattering and
obeys the usual Bose thermal population factor, whereas the 
quasi-elastic intensity increases with decreasing temperature,
but then collapses below $T_C$. If we subtract an elastic
nuclear incoherent contribution measured at 10 K, we see that the change
at $T_C$ is entirely due to the quasi-elastic scattering contribution,
showing that the lattice distortions giving rise to it are
quasi-static on a time scale $\tau \sim \hbar /2\Delta E\sim $ 1 ps set by
the energy resolution of the instrument, i.e. they are static on
the time scale of typical phonon vibrations. A good description of 
the {\bf q}-dependence of this diffuse scattering can be
obtained in terms of Huang scattering, consistent with a Jahn-Teller 
type distortion around the Mn$^{3+}$ ions \cite{preosti}. Our 
results therefore provide direct evidence both for the existence of 
quasi-static polarons above $T_C$, and their abrupt disappearance upon 
cooling below the ferromagnetic 
transition, where the charges delocalize \cite{mitchell,prl}.

The measurements also reveal the presence of broad incommensurate
peaks in the paramagnetic phase, as shown by the contour 
plot of the x-ray intensity at 125 K in the ($hk$) plane at $l$=18 
in Fig. 2(a). Three broad peaks are observed around the diffuse 
scattering rod; the expected fourth peak was not experimentally 
accessible. These peaks are characterized by a wave vector 
($\pm \epsilon $, 0, $\pm 1$) as measured from the nearest 
fundamental Bragg peak, where $\epsilon \simeq 0.3$ (in terms of 
reciprocal lattice units 
($2\pi /a$, $0$, $2\pi /c$)). (0, $\pm \epsilon $, $\pm 1$) peaks 
are also observed, either because of the presence of ($a$,$b$) twin 
domains in a 1{\bf q}-system, or because this is a 2{\bf q}-system.
The in-plane incommensurability is evident in the x-ray $h$-scans 
shown in Fig. 2(b) at different temperatures.
Note that, similar to the quasi-elastic peak in Fig. 1(c), this peak
increases and then rapidly decreases in intensity as we cool through
$T_C$. Figure 2(c) shows various neutron scans along the $l$-direction 
through the incommensurate peak positions (2.3, 0, $l$). The red circles 
are for an energy-integrated scan that reveals two broad symmetric 
peaks at $l = \pm 1$, consistent with out-of-phase correlations between 
bilayers. Identical data are obtained in (energy integrated) x-ray 
scans. The orange circles depict an elastic neutron scan across one of 
the peaks, scaled by an instrumental factor. Energy scans at the peak 
positions have confirmed that the correlations 
giving rise to these peaks are once again quasi-static on a
time scale $\tau \sim $ 1 ps. We 

\begin{figure}
\vspace*{-3.30cm}
\centerline{\epsfig{file=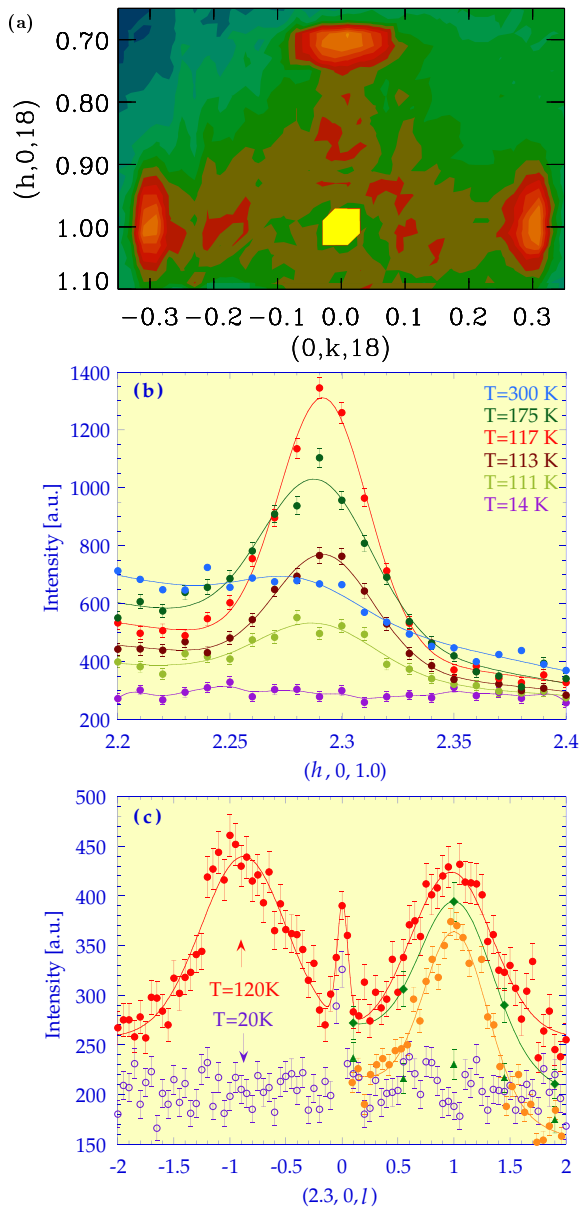,width=15.2truecm}}

\vspace*{-2.4cm}
\caption{Polaron ordering in La$_{1.2}$Sr$_{1.8}$Mn$_2$O$_7$. (a)
Contour plot of the x-ray intensity in the ($hk$) plane at $l=18$,
collected at $T$=125 K. Three incommensurate peaks
are observed, characterized by the wave vector ($\epsilon$, 0, $l$)
or (0, $\epsilon$, $l$). The expected fourth
peak was not accessible experimentally. The intensity at (1, 0, 18)
is due to the rod of scattering from stacking faults
crossing this plane. An absorption correction
based on the rotation of the sample has been applied to these data. 
(b) X-ray $h$-scans through the incommensurate peak (2.3, 0, 1)
at different temperatures. The higher scattering at small $h$ is due to
the proximity of the lobe-shaped diffuse scattering around the Bragg
peak. (c) Neutron $l$-scans through the charge ordering peaks at
(2.3, 0, $\pm$1) at $T$=120 K: energy-integrated (red), elastic
(orange), non-spin-flip scattering measured with polarized neutrons 
(green diamonds), spin-flip scattering (green triangles). The elastic 
and non-spin-flip data points have been scaled by appropriate 
instrumental factors. The $l$-scan at $T$=20 K (open
purple circles) shows that the charge ordering peaks have vanished.}
\end{figure}

\noindent
have also performed polarized neutron experiments to 
probe the nature of this scattering.
In the configuration where the neutron polarization {\bf P} 
$\parallel $ {\bf Q} (the neutron wave vector), we found
that all the signal was non-spin-flip scattering (depicted by the 
green diamonds in Fig. 2(c)), while any magnetic scattering would be
spin-flip (green triangles in Fig. 2(c)). Thus the incommensurate
peaks are purely structural reflections.

Figure 3(a) shows that the temperature dependence of the 
incommensurate peak intensity is remarkably similar to the Huang 
scattering, whether the latter is derived from the x-ray scattering by 
subtracting the estimated thermal diffuse scattering (straight line in
Fig. 1(c)) or directly from the quasi-elastic neutron scattering. This 
indicates that both types of scattering are associated with the 
development of polarons above $T_C$. 
The incommensurate peak intensity falls slightly
more rapidly than the Huang scattering with increasing temperature.
This is consistent with ascribing the Huang scattering to individual
polarons, and the incommensurate peaks to polaron correlations which
become stronger with decreasing temperature.
Below $T_C$ we observe a ``melting" of the polaron 
correlations occurring simultaneously with the collapse of the polarons 
themselves. We note that the collapse of the polaron correlations also
occurs under an applied magnetic field (see Fig. 3(b)). This behavior 
is expected because of the coupling between the charge and spin 
dynamics through the double exchange interaction.

The incommensurate peaks are broader than the $q$ resolution,
showing that the in-plane and out-of-plane charge correlations remain 
relatively short range at all temperatures. Detailed measurements in 
the ($h,0,l$) plane indicate that the correlation lengths are weakly
temperature dependent and peak at the same temperature as the intensity, 
with $\sim 26.4$ \AA\ $\sim 6a$ in-plane, and 
$\sim 10.4$ \AA\ $\sim \frac{c}{2}$ out-of-plane. No higher harmonics 
have been observed, and no superlattice peaks have been found in the 
($hhl$) plane. The $l = \pm 1$ component of the charge ordering 
wave vector is related to the presence of two MnO$_{2}$ bilayers per 
unit cell, and indicates that distortions produced by the modulation 
of the charge density are, on average, out of phase in adjacent 
bilayers. This results in a staggering of the charges from one bilayer 
to the next, as would be expected from Coulomb repulsion. The small 
$c$-axis correlation length ($\simeq$ the separation between two 
bilayers) suggests that only two bilayers are correlated at most.
The short-range nature of the charge correlations makes it difficult to
collect enough integrated intensities at this stage to perform 
quantitative comparisons to specific charge ordering models.

Charge and orbital ordering have been observed at low temperature in a
number of insulating, antiferromagnetic cubic manganites at small 
\cite{murakami,hirota} and large ($x\geq 0.5$) \cite{cheong} doping, 
as well as in layered manganites with $x$=$0.5$ \cite{li}.
Commensurate charge modulations in the antiferromag-

\begin{figure}
\vspace*{-3.65cm}
\centerline{\epsfig{file=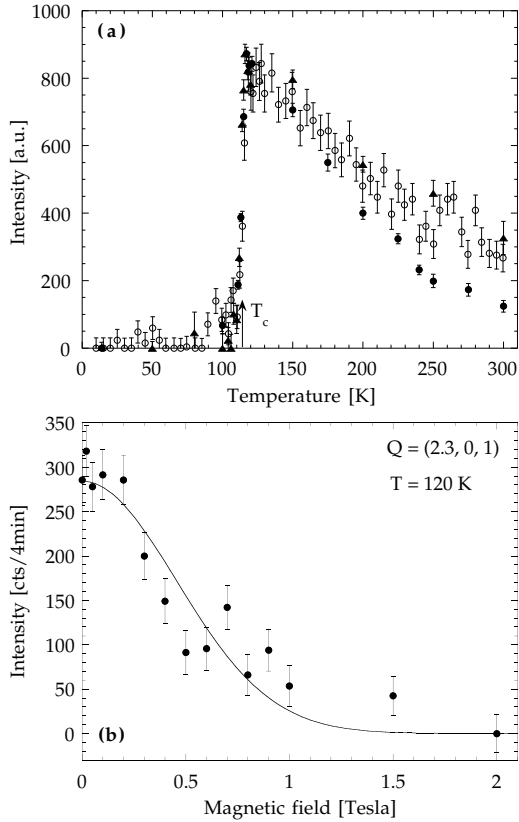,width=12.3truecm}}

\vspace*{-2.8cm}
\caption{(a) Temperature-dependence of the x-ray intensity of the
(2.3, 0, 1) incommensurate peak (closed circles), of the diffuse x-ray
scattering after correction for the phonon contribution (closed
triangles), and of the quasi-elastic neutron peak (open circles)
in Fig. 1(c). The diffuse scattering due to the strain field around
the localized charges (polarons) and the satelite peaks due to polaron
ordering collapse together at $T_C$. (b) Field-dependence
of the intensity of the (2.3, 0, 1) peak at $T$=120 K.}
\end{figure}

\noindent
netic insulating 
phases are also a familiar scenario in the related nickel oxides 
\cite{chen}. However, short-range charge ordering in the paramagnetic 
phase of an optimally-doped CMR ferromagnet is a novel feature
observed here. The charge correlations result from
Coulomb interactions between the polarons, coupled with the interaction 
of overlapping polaronic strain fields. In the present $x$=$0.4$ system 
they are not strong enough to win the competition with the double 
exchange interaction, and the charges delocalize at the ferromagnetic 
transition, where the charge peaks collapse and the lattice strain 
relaxes. It is the delicate balance between double exchange, Coulomb 
repulsion and the lattice strain field that dictates whether the 
material is a ferromagnetic metal or charge-ordered insulator at 
low temperatures.

This work was supported by the U.S. Department of Energy,
Basic Energy Sciences-Materials Sciences (W-31-109-ENG-38),
NSF (DMR 97-01339), NSF-MRSEC (DMR 96-32521), and 
the Swiss National Science Foundation. We are grateful to N. 
Wakabayashi for drawing our attention to similar x-ray diffuse 
scattering observations \cite{shimomura}, and
to W.-K. Lee for his help with the x-ray measurements.


\begin{references}
\bibitem{magnrod} The sharp rod of scattering along the [0, 0, $l$]
direction is resolution limited in the ($h$$k$0) plane and is associated
with stacking faults inevitably found in crystals with layered
structures.
\end{references}
\end{document}